# A Comparative Evaluation of CT Global Noise Calculation Methods for Clinical Image Quality Assessment

Charles M Weaver, MS, Gary Ge, PhD, Alexander Alsalihi, Jie Zhang, PhD

Division of Diagnostic and Nuclear Medical Physics, Department of Radiology, University of Kentucky College of Medicine, Lexington, KY 40536

## Abstract

**Background:** The recently introduced CMS quality measure for computed tomography (CT) requires compliance with two key metrics, radiation dose and image quality, across 18 exam categories. However, the method for calculating global noise (GN) remains undefined, with references to the "Duke method" and "Wisconsin method" as possible options. This lack of clarity raises concerns about standardization, compliance, and clinical relevance.

**Purpose:** To compare GN calculation methods proposed in the Duke and Wisconsin papers and evaluate their variability and reliability across clinical CT protocols.

**Methods:** A retrospective analysis of 719 CT exams was conducted across seven protocols, including five abdominal and two chest categories. One protocol (Chest PE) included exams reconstructed with both smooth and sharp kernels. Five GN metrics were evaluated: Duke_tissue_mean, Wisconsin_tissue_mean, Wisconsin_tissue_mode, Wisconsin_air_mean, and Wisconsin_air_mode. Statistical differences were assessed using the Friedman test with pairwise Wilcoxon signed-rank tests, and Pearson correlation matrices were used to evaluate agreement.

**Results:** Significant differences ($p < 0.05$) were observed across GN metrics in all protocols, with Wisconsin_tissue_mean consistently producing the highest values. Correlation analysis showed strong agreement ($r > 0.7$) for Renal Stone, Chest, and Abdomen/Pelvis protocols, but weaker correlations for Urogram, Renal Mass, and Enterography. Mode-based metrics showed higher agreement (r > 0.9), suggesting dose dependency. In the Chest PE protocol, the smooth kernel yielded GN values well below the CMS threshold, while the sharp kernel exceeded the threshold in tissue-based metrics.

**Conclusions:** Significant variability across GN metrics highlights the need for a standardized, clinically relevant method. Without clear definitions, the CMS measure's effectiveness in ensuring image quality and dose management may be compromised, an issue also raised by the AAPM-commissioned panel.

# Introduction

The Centers for Medicare & Medicaid Services (CMS) have implemented a new CT quality measure designed to balance radiation dose reduction with the maintenance of diagnostic image quality[1]. This measure is used to assess the proportion of studies that surpass established thresholds for image noise or radiation dose across various CT examination categories. Integrated into major CMS quality-based payment programs, this initiative influences reimbursements for hospitals and clinics, with reporting commencing in January 2025 and mandatory compliance slated for 2027.

A critical component of this measure is the calculation of CT global noise (GN), a metric indicative of image quality. However, CMS does not prescribe a specific methodology for this calculation, instead referencing two existing methods developed by Wisconsin and Duke universities, respectively [2-4]. The Wisconsin method employs an automated algorithm to measure noise in CT images, generating a GN level that reflects the standard deviation of Hounsfield units within specified regions of interest. Conversely, the Duke method focuses on calculating GN magnitude by analyzing soft tissue regions in patient CT images.

The availability of two different methods provides flexibility in reporting, but simultaneously introduces potential variability when assessing CT image quality across institutions. This possibility would lead to GN calculation dependencies and compromise the efficacy of the CMS measure in evaluating institutional image quality. This study aims to critically assess two existing methods for GN calculations, examining their application across various clinical CT protocols to identify potential discrepancies and implications for maintaining diagnostic integrity while adhering to radiation dose optimization goals.

## Methods

*Global Noise Calculation*

Five noise-related metrics were derived from two established methods: the Duke method and the Wisconsin method. The Wisconsin method proposed performing HU measurements in air or soft tissue and calculating the noise metric using the mean or mode of the measurements, resulting in four possible permutations for GN[3]. Accordingly, the five GN metrics were named as follows: Duke_tissue_mean, Wisconsin_tissue_mean, Wisconsin_tissue_mode, Wisconsin_air_mean, Wisconsin_air_mode.

Implementation of the GN metrics was done according to the published methodology. Metrics were generated on a slice-by-slice basis and averaged across slices to produce GN values per patient. Because the Wisconsin method specified the use of 7mm regions of interest (ROI), a 7mm sliding window was used for the Duke method to maintain consistency. While this differs slightly from the Duke study, which used a 6mm window, the study demonstrated that window sizes between 6mm and 20mm produced consistent results[2]. Calculations for all metrics were done using Python 3.12 (Python Software Foundation, www.python.org).

Patient images for seven CT protocols were retrospectively retrieved for GN metric evaluation, which included five abdomen and two chest protocols. The abdomen protocols included Renal Stone (Low Dose), Abdomen/Pelvis (Routine Dose), Enterography (Routine Dose), Urogram

(High Dose), and Renal Mass WO (High Dose). The chest protocols included Chest WO (Routine Dose) and Chest PE (Routine Dose). The selection criteria for images was that all metrics were able to be calculated for each exam (i.e., air was present in the images).

The protocols were chosen to cover a range of CT exam categories based on examples outlined in the UCSF study referenced in the CMS measure[5]. Selecting multiple protocols within a single category (e.g. Abdomen/Pelvis and Enterography) allowed for intra-category assessments of the GN metrics.

A total of 719 exams were collected for the study cohort, with at least 100 exams per protocol. For the Chest PE protocol, patients were split evenly between a smooth and sharp reconstruction kernel to illustrate the impact of kernel selection on GN metrics. This retrospective study was performed in accordance with relevant guidelines and regulations and was approved by the University of Kentucky institutional review board (IRB).

*Statistical Analysis*

Friedman tests were run between all metrics for each CT protocol to evaluate differences between GN metrics. Subsequent pairwise Wilcoxon signed-rank tests were used to determine which pairs of metrics were statistically different. Pearson correlation matrices were generated to assess agreement between GN metrics across protocols. Additionally, paired permutation tests assessed whether the GN metrics were significantly different across individual exams.

**Results**

Wisconsin_tissue_mean generates the largest GN values and Wisconsin_air_mode generates the smallest GN values consistently. Duke_tissue_mean generates the lowest GN value among the tissue-based metrics for all protocols except Chest PE. Friedman tests and subsequent pairwise analysis demonstrates significant differences ($p < 0.05$) between GN metrics in all protocols (Figure 1). Paired permutation tests revealed significant differences in tissue noise metrics for nearly all patients, with only five out of 719 cases showing no significant variation. Larger variability was noted in the Urogram protocol compared to other exam types, indicating potential limitations in the reliability of certain noise metrics for high-dose studies.

Correlation analysis indicated strong agreement ($r > 0.7$) among the five global noise metrics for Renal Stone, Chest, and Abdomen/Pelvis protocols. Weaker correlations were observed for Urogram, Renal Mass, and Enterography protocols. The strength of these correlations appeared to be dose-dependent, with mode-based global noise metrics exhibiting stronger agreement, often exceeding $r > 0.9$.

Correlation matrices (Figure 2) illustrate that the Abdomen/Pelvis protocol demonstrated strong consistency, with eight out of ten comparisons showing very strong correlations ($r \geq 0.8$) and the remaining two classified as strong ($0.6 \leq r < 0.8$). Similarly, the Chest protocol exhibited nine very strong correlations ($r \geq 0.8$) and one strong correlation ($0.6 \leq r < 0.8$). The Renal Stone protocol showed the highest level of agreement, with all ten comparisons classified as very strong ($r \geq 0.8$). In contrast, the Urogram, Renal Mass, and Enterography protocols displayed weak correlations, highlighting inconsistencies in GN metric performance across different CT exam types.

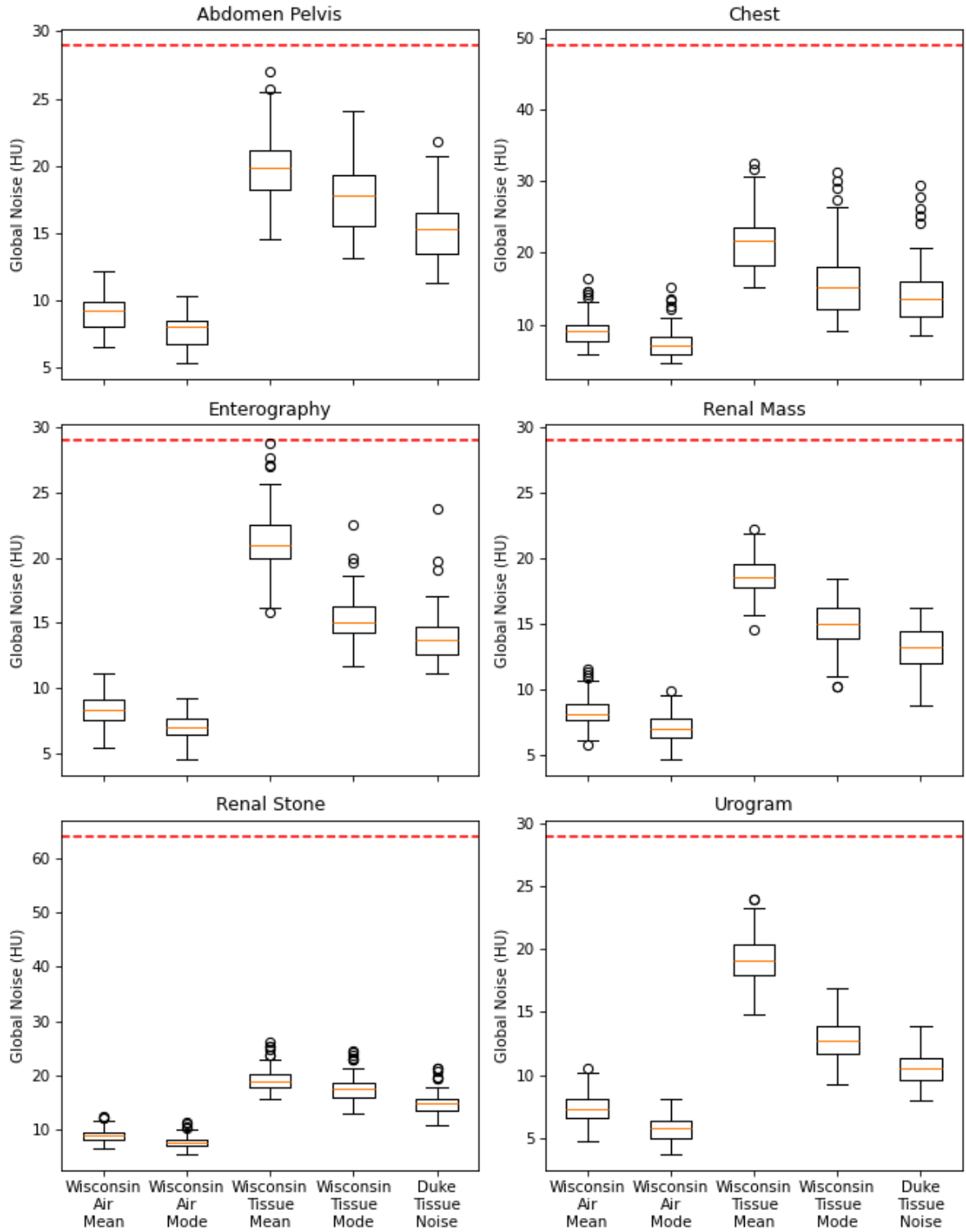

**Figure 1**: Box plots of Global Noise (GN) metrics for six protocols, highlighting differences between Wisconsin_tissue_mean, Wisconsin_tissue_mode, and Duke_tissue_mean for Renal Mass, Chest, Abdomen/Pelvis, Enterography, Renal Stone, and Urogram protocols. The air-based metrics generate lower GN values than tissue-based metrics, and the Wisconsin_tissue_mean metric is consistently the highest. All metrics are significantly different from each other ($p<0.05$), indicating that metric selection will impact an institution's reported values.

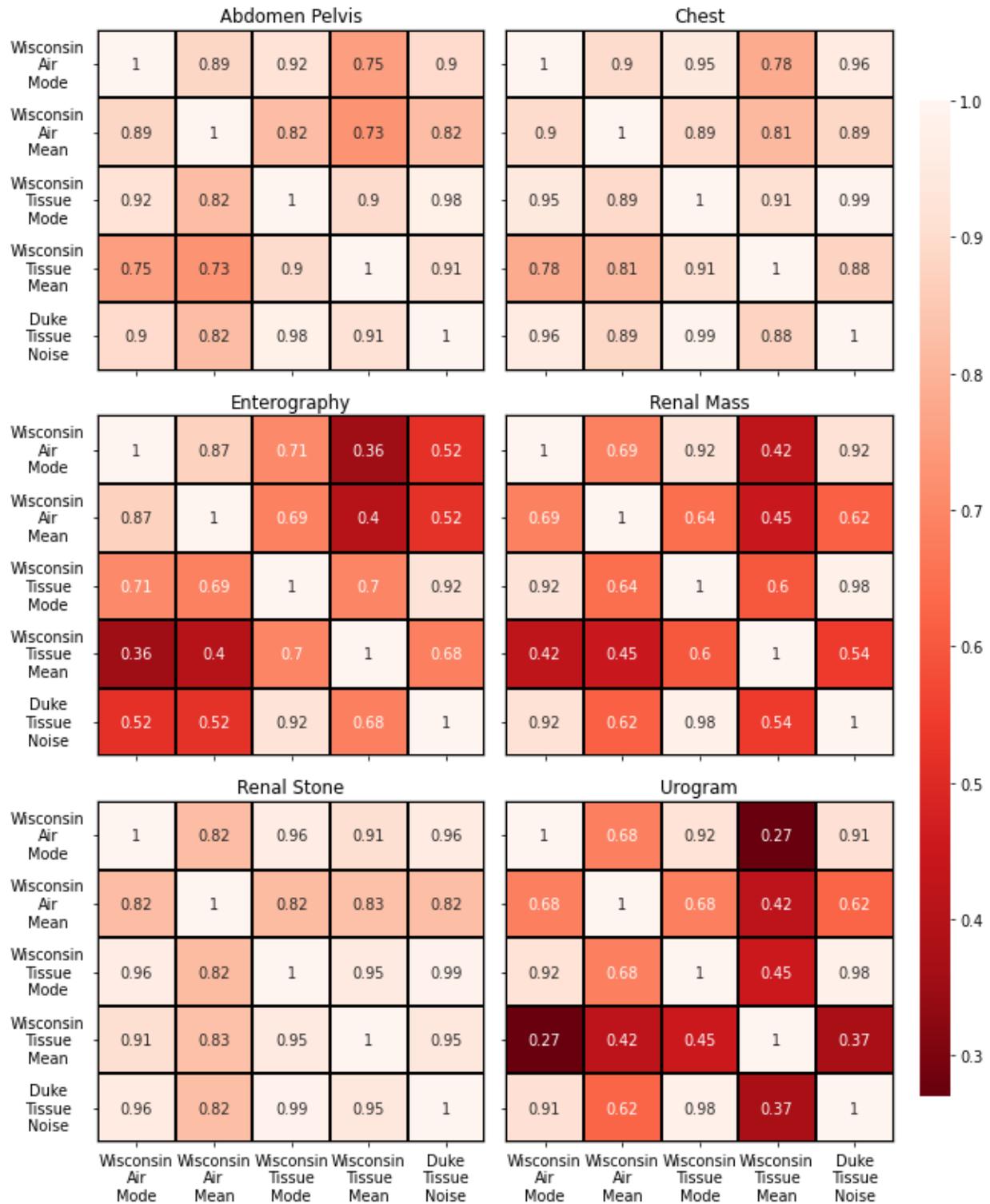

**Figure 2**: Heatmaps showing correlation strengths for global noise metrics per protocol. Abdomen/Pelvis: 8/10 show very strong correlations (r ≥ 0.8), 2/10 strong (0.6 ≤ r < 0.8); Chest: 9/10 very strong, 1/10 strong; Renal Stone: 10/10 very strong, Urogram: 3/10 very strong, 3/10 strong, 2/10 moderate (0.4 ≤ r < 0.6), 2/10 weak (0.2 ≤ r < 0.4); Enterography: 2/10 very strong, 4/10 strong, 3/10 moderate, 1/10 weak; Renal Mass: 3/10 very strong, 4/10 strong, 3/10 moderate.

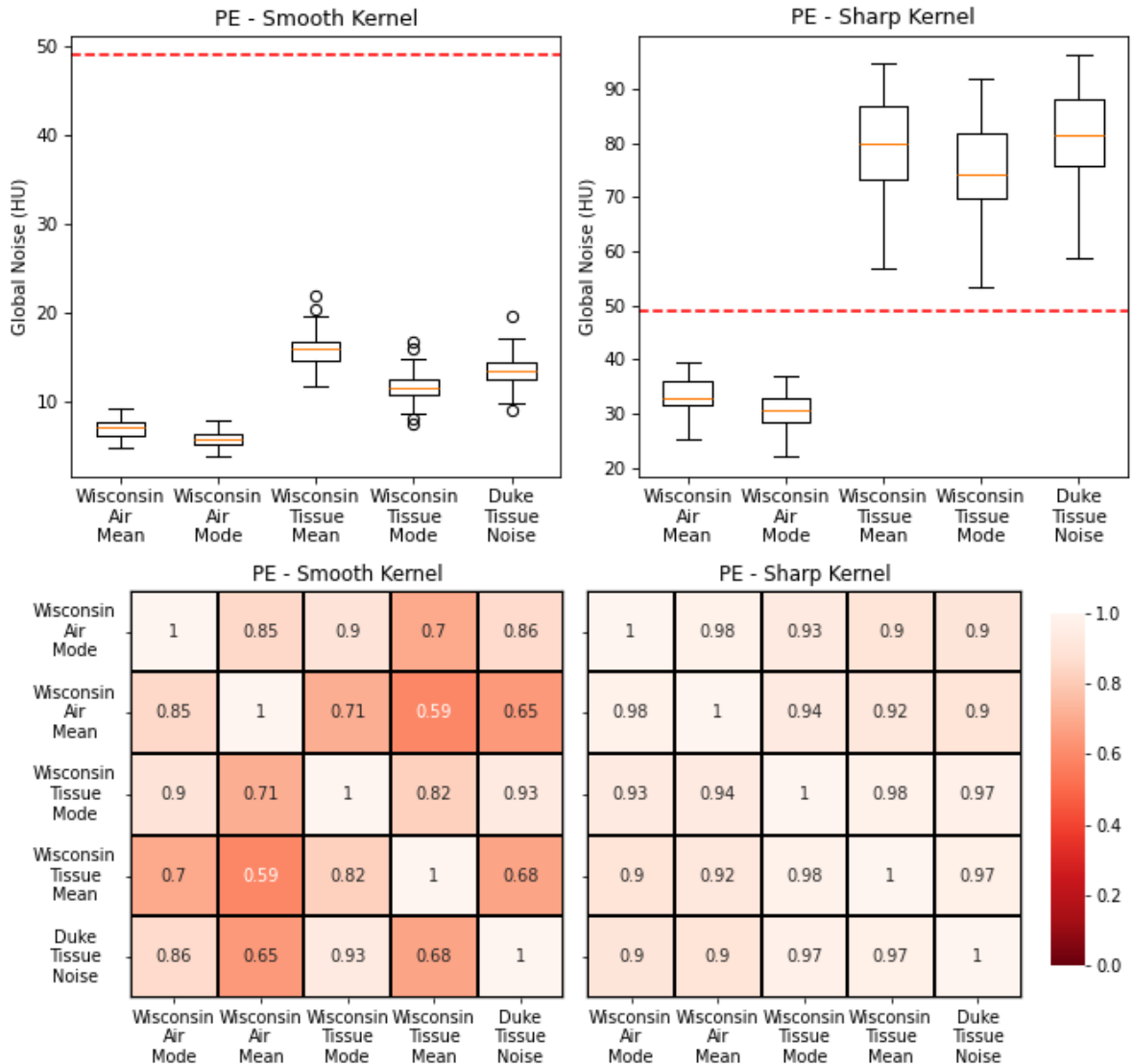

**Figure 3**. Boxplots and correlation matrices for the Chest PE protocol. The patient cohort was evenly split between a smooth and sharp reconstruction kernel. The global noise values for the sharp kernel exceed the CMS threshold in the tissue noise metrics. The correlation matrix for the smooth kernel shows low correlations similar to the Urogram protocol, where the sharp kernel has high correlations similar to the Chest WO protocol.

The results for the Chest PE protocol are shown in Figure 3. The smooth reconstruction kernel has GN metric values that are well below the CMS-defined threshold, where the sharp reconstruction kernel has values that are above the threshold in the tissue noise metrics. The correlation matrix for the smooth reconstruction kernel shows similar behavior to the Urogram (High Dose) and Renal Mass (High Dose) CT protocols, whereas the sharp kernel is similar to the Renal Stone (Low Dose), Chest (Routine dose) and Abdomen/Pelvis (Routine Dose) protocols.

## Discussion

We observe significant variations across the five GN metrics when applied to different CT protocols. Each metric also behaves differently across protocols, which may be due to differences in the examinations and scan parameters. While some variation is expected, GN values vary substantially based on the chosen metric, with air-based metrics consistently yielding lower values. Even among similar metric types, notable discrepancies are present.

GN is highly sensitive to variables such as reconstruction algorithms, iterative reconstruction (IR) strength, and kernel selection. However, standardizing CT protocols is nearly impossible due to institutional dependencies (e.g., radiologist preferences, local patient populations, vendor selections) and both inter- and intra-vendor variability. For example, the AAPM reference lung cancer screening protocols shows variation in scan parameters between vendors as well as different models made by the same vendor. Modern image reconstruction methods (e.g., IR, deep-learning) add additional complexity to this through different reconstruction strengths. Although CMS addresses slice thickness by standardizing to a 3 mm reference, it does not account for other scan parameters. For example, as shown in Figure 3, Chest PE protocols typically use sharp kernels for clinical diagnosis, but a smooth reconstruction that is part of the protocol would result in lower image noise and would influence GN values, potentially misrepresenting true clinical image quality and undermining the intent of the CMS measure. Furthermore, focusing only on global noise amplitude ignores noise texture, which significantly impacts perceived image quality and diagnostic confidence.

It is worth noting that the current CMS measure lacks clarity regarding both the choice of GN metric and the method for defining GN thresholds. This ambiguity may enable institutions to report artificially low GN values that fall below thresholds, irrespective of actual image quality. However, the National Quality Forum (NQF) submission for the CMS measure states [4] "Noise as defined in this measure is calculated on every CT image within a scan (a single irradiating event), and the global noise value for each scan is the mean value across all images. For CT exams that have multiple scans…the exam is assigned the 'best' global noise value across all scans, i.e., the highest quality scan." Based on this statement, the NQF submission implies that all reconstructed image series from an irradiating event are included in GN calculations, regardless of their clinical relevance. This approach introduces additional variability and creates opportunities for gaming, particularly when non-diagnostic series are present.

The selection of the GN threshold requires reconsideration. Our findings suggest that current thresholds for certain protocols, such as Renal Stone and Chest WO, may be too high. Threshold selection should be guided by the image quality metric used, where tissue-based metrics require higher thresholds, and air-based metrics require lower ones. However, variability in acquisition techniques and the lack of standardized reconstruction parameters or noise metrics make establishing meaningful thresholds a significant challenge.

Our findings align with concerns raised by the AAPM-commissioned panel evaluating the CMS measure [6]. Without a standardized GN metric and reporting framework, the measure may fail to achieve its intended goals of promoting consistent CT image quality and dose optimization.

The choice of GN metric directly impacts compliance and inter-institutional comparability. A clearly defined, clinically meaningful, and universally adopted metric is essential to ensure: consistent application of regulatory standards; reliable image quality assessment, and fair comparison of dose optimization efforts across institutions.

This study has several limitations. We did not evaluate all 18 CMS-listed protocols, but this will not impact the validity of our main conclusions. Due to the retrospective nature of the study, we also did not investigate the impact of slice thicknesses other than 3mm or IR reconstruction strength. However, the significant differences observed between two reconstruction kernels in the PE protocol suggest that similar variability would be seen with other acquisition parameters. In addition, the selection criteria for the patient cohort ensured that both the air- and tissue-based metrics were able to be calculated, but it is possible that CT protocols could acquire images without air (e.g., small FOV cardiac studies, large patient habitus). How these cases would be handled does not appear to be clarified in the measure document.

In conclusion, this study provides a critical assessment of two methods for calculating CT global noise referenced in the upcoming CMS measure. By comparing five GN metrics across multiple clinical protocols, we identify substantial metric-dependent variability. These findings underscore the need for a standardized, clinically appropriate GN metric to ensure consistency in image quality evaluation and compliance with regulatory standards.

## Citations

1. *Excessive Radiation Dose or Inadequate Image Quality for Diagnostic Computed Tomography (CT) in Adults*. 2024, Centers for Medicare and Medicaid Services.

2. Christianson, O., et al., *Automated Technique to Measure Noise in Clinical CT Examinations*. AJR Am J Roentgenol, 2015. **205**(1): p. W93-9.
3. Malkus, A. and T.P. Szczykutowicz, *A method to extract image noise level from patient images in CT*. Med Phys, 2017. **44**(6): p. 2173-2184.
4. *Excessive Radiation Dose or Inadequate Image Quality for Diagnostic Computed Tomography (CT) in Adults (Clinician Level)*. 2022, National Quality Forum.
5. Smith-Bindman, R., et al., *An Image Quality-informed Framework for CT Characterization*. Radiology, 2022. **302**(2): p. 380-389.
6. Wells, J.R., et al., *The New CMS Measure of Excessive Radiation Dose or Inadequate Image Quality in CT: Issues and Ambiguities—Perspectives from an AAPM-Commissioned Panel*. American Journal of Roentgenology, 2025.